\documentclass[12pt,a4paper,color]{article}
\usepackage{showkeys}
\usepackage{epsfig}
\usepackage{fullpage}
\usepackage{algorithm}
\usepackage{graphicx}
\usepackage{amssymb}
\usepackage{amsmath}
\usepackage{axodraw}
\usepackage{tikz}
\usepackage{datetime}

\begin{document}

\title{Artificial spin-ice and vertex models}
\author{Leticia F. Cugliandolo \\
{\small Sorbonne Universit\'es, Universit\'e Pierre et Marie Curie,} \\
{\small Laboratoire de Physique Th\'eorique et Hautes Energies,} \\
{\small Tour 13, 5\`eme \'etage, 4 Place Jussieu, } \\
{\small 75005 Paris France}
}

\maketitle

\abstract{
In classical and quantum frustrated magnets  the interactions in combination with the lattice structure impede the spins to order in 
optimal configurations at zero temperature. The theoretical interest in their classical realisations has been boosted 
by the artificial manufacture of materials with these properties, that are of flexible design. 
This note summarises work on the use of vertex models to study bidimensional spin-ices samples, 
done in collaboration with 
R. A. Borzi, M. V. Ferreyra, L. Foini, G. Gonnella, S. A. Grigera, P. Guruciaga,
D. Levis,  A. Pelizzola and M. Tarzia, in recent years. It 
is an invited contribution to a J. Stat. Phys. special issue dedicated to the memory of Leo P. Kadanoff. 
%\\
%$\;$
%\\
%\today\qquad\currenttime
}

\newpage

\section{Introduction}

Vertex models were introduced to describe phase transitions in 
ferro-electric systems. Their analysis needed the use of sophisticated tools of mathematical physics and 
motivated the development of many fancy methods since the 70s~\cite{BaxterBook,Wuphasetransitions,Reshetikhin}. 

Very recently, vertex models have been used to 
model the statics and dynamics of 2D artificial spin samples~\cite{Marrows16}. This is the aspect of these models that I will 
dwell upon in this note, that is organised as follows.
After an introduction to vertex models, and their mapping to 
a model with multi-spin interactions introduced by L. Kadanoff and others, I will explain what artificial spin-ice samples are. I will then 
give some guidelines on the approach adopted 
and the results found in a number of works devoted to the use of vertex model to better understand the 
behaviour of these artificial magnets~\cite{Demian-thesis,Levis2013a,Foini2013,Levis2012,Levis2013b,CuGoPe,Guruciaga16}.
The paper ends with a short conclusion.

\section{Vertex models}

We start by recalling the definition, and a number of very well-known properties, of vertex models in two dimensions.

\vspace{0.25cm}
\noindent
{\bf The six vertex model}
\vspace{0.25cm}

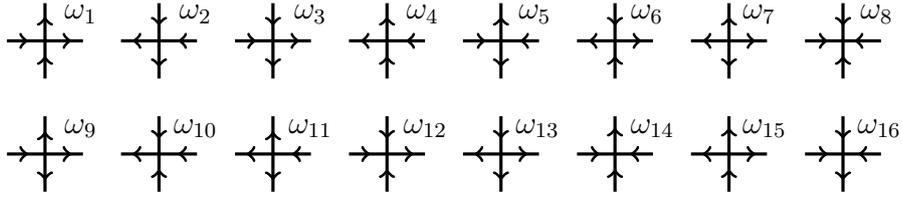
\begin{figure}[h]
\begin{center}
\begin{tikzpicture}
% vertex a1 
   \draw [very thick, -] (-8,0) -- (-7,0);
   \draw [very thick, -] (-7.5,-0.5) -- (-7.5,0.5);
   \Text(-200,10)[]{$\omega_1$}; 
   % arrows
   \draw [very thick, ->](-7.2,0) -- (-7.15,0);
   \draw [very thick, ->](-7.75,0) -- (-7.72,0);
   \draw [very thick, ->] (-7.5,-0.25) -- (-7.5,-0.2);
   \draw [very thick, ->] (-7.5,0.27) -- (-7.5,0.32);
% vertex a2
   \draw [very thick, -] (-6.5,0) -- (-5.5,0);
   \draw [very thick, -] (-6,-0.5) -- (-6,0.5);
   \Text(-157,10)[]{$\omega_2$}; 
   % arrows
   \draw [very thick, ->](-6.3,0) -- (-6.35,0);
   \draw [very thick, ->](-5.72,0) -- (-5.75,0);
    \draw [very thick, ->] (-6,-0.3) -- (-6,-0.35);
   \draw [very thick, ->] (-6,0.25) -- (-6,0.2);
% vertex b1
   \draw [very thick, -] (-5,0) -- (-4,0);
   \draw [very thick, -] (-4.5,-0.5) -- (-4.5,0.5);
   \Text(-114,10)[]{$\omega_3$}; 
   % arrows
   \draw [very thick, ->](-4.2,0) -- (-4.15,0);
   \draw [very thick, ->](-4.75,0) -- (-4.72,0);
   \draw [very thick, ->] (-4.5,-0.3) -- (-4.5,-0.35);
   \draw [very thick, ->] (-4.5,0.25) -- (-4.5,0.2);
   % vertex b2 
   \draw [very thick, -] (-3.5,0) -- (-2.5,0);
   \draw [very thick, -] (-3,-0.5) -- (-3,0.5);
   \Text(-72,10)[]{$\omega_4$}; 
   % arrows
   \draw [very thick, ->](-3.3,0) -- (-3.35,0);
   \draw [very thick, ->](-2.72,0) -- (-2.75,0);
   \draw [very thick, ->] (-3,-0.25) -- (-3,-0.2);
   \draw [very thick, ->] (-3,0.27) -- (-3,0.32);
% vertex c1
   \draw [very thick, -] (-2,0) -- (-1,0);
   \draw [very thick, -] (-1.5,-0.5) -- (-1.5,0.5);
   \Text(-30,10)[]{$\omega_5$}; 
   %arrows
   \draw [very thick, ->] (-1.5,-0.3) -- (-1.5,-0.35);
   \draw [very thick, ->] (-1.5,0.27) -- (-1.5,0.32);
   \draw [very thick, ->](-1.75,0) -- (-1.7,0);
   \draw [very thick, ->](-1.2,0) -- (-1.25,0);
% vertex c2
   \draw [very thick, -] (-0.5,0) -- (0.5,0);
   \draw [very thick, -] (0,-0.5) -- (0,0.5);
   \Text(12,10)[]{$\omega_6$}; 
   %arrows
   \draw [very thick, ->] (0,-0.25) -- (0,-0.2);
   \draw [very thick, ->] (0,0.25) -- (0,0.2);
   \draw [very thick, ->](0.3,0) -- (0.35,0);
  \draw [very thick, ->](-0.3,0) -- (-0.35,0);
   % vertex d1
   \draw [very thick, -] (1,0) -- (2,0);
   \draw [very thick, -] (1.5,-0.5) -- (1.5,0.5);
   \Text(54,10)[]{$\omega_7$}; 
   %arrows
   \draw [very thick, ->] (1.5,-0.3) -- (1.5,-0.35);
   \draw [very thick, ->] (1.5,0.27) -- (1.5,0.32);
   \draw [very thick, ->](1.2,0) -- (1.15,0);
   \draw [very thick, ->](1.8,0) -- (1.85,0);
% vertex d2
   \draw [very thick, -] (2.5,0) -- (3.5,0);
   \draw [very thick, -] (3,-0.5) -- (3,0.5);
   \Text(98,10)[]{$\omega_8$}; 
   %arrows
   \draw [very thick, ->] (3,-0.25) -- (3,-0.2);
   \draw [very thick, ->] (3,0.25) -- (3,0.2);
   \draw [very thick, ->](2.75,0) -- (2.8,0);
   \draw [very thick, ->](3.2,0) -- (3.15,0);
%%%%%%%%%%%%%%%%%%%%%%%
% vertex e1 
   \draw [very thick, -] (-8,-1.5) -- (-7,-1.5);
   \draw [very thick, -] (-7.5,-2) -- (-7.5,-1);
   \Text(-200,-33)[]{$\omega_9$}; 
   % arrows
   \draw [very thick, ->](-7.2,-1.5) -- (-7.15,-1.5);
   \draw [very thick, ->](-7.75,-1.5) -- (-7.72,-1.5);
   \draw [very thick, ->] (-7.5,-1.25) -- (-7.5,-1.2);
   \draw [very thick, ->] (-7.5,-1.8) -- (-7.5,-1.85);
% vertex e2
   \draw [very thick, -] (-6.5,-1.5) -- (-5.5,-1.5);
   \draw [very thick, -] (-6,-2) -- (-6,-1);
   \Text(-157,-33)[]{$\omega_{10}$}; 
   % arrows
   \draw [very thick, ->](-5.7,-1.5) -- (-5.75,-1.5);
   \draw [very thick, ->](-6.3,-1.5) -- (-6.35,-1.5);
   \draw [very thick, ->] (-6,-1.25) -- (-6,-1.3);
   \draw [very thick, ->] (-6,-1.75) -- (-6,-1.7);
% vertex e3
   \draw [very thick, -] (-5,-1.5) -- (-4,-1.5);
   \draw [very thick, -] (-4.5,-2) -- (-4.5,-1);
   \Text(-114,-33)[]{$\omega_{11}$};
   % arrows 
   \draw [very thick, ->] (-4.5,-1.25) -- (-4.5,-1.2);
   \draw [very thick, ->] (-4.5,-1.8) -- (-4.5,-1.85);
   \draw [very thick, ->](-4.15,-1.5) -- (-4.25,-1.5);
   \draw [very thick, ->](-4.8,-1.5) -- (-4.85,-1.5);
   % vertex e4 
   \draw [very thick, -] (-3.5,-1.5) -- (-2.5,-1.5);
   \draw [very thick, -] (-3,-2) -- (-3,-1);
   \Text(-71,-33)[]{$\omega_{12}$}; 
   % arrows
   \draw [very thick, ->] (-3,-1.25) -- (-3,-1.3);
   \draw [very thick, ->] (-3,-1.75) -- (-3,-1.7);
   \draw [very thick, ->](-2.7,-1.5) -- (-2.65,-1.5);
   \draw [very thick, ->](-3.25,-1.5) -- (-3.22,-1.5);
% vertex e5
   \draw [very thick, -] (-2,-1.5) -- (-1,-1.5);
   \draw [very thick, -] (-1.5,-2) -- (-1.5,-1);
    \Text(-29,-33)[]{$\omega_{13}$}; 
   % arrows
    \draw [very thick, ->] (-1.5,-1.25) -- (-1.5,-1.3);
    \draw [very thick, ->] (-1.5,-1.8) -- (-1.5,-1.85);
    \draw [very thick, ->](-1.2,-1.5) -- (-1.15,-1.5);
    \draw [very thick, ->](-1.82,-1.5) -- (-1.85,-1.5);
% vertex e6
   \draw [very thick, -] (-0.5,-1.5) -- (0.5,-1.5);
   \draw [very thick, -] (0,-2) -- (0,-1);
   \Text(14,-33)[]{$\omega_{14}$}; 
   % arrows
    \draw [very thick, ->] (0,-1.2) -- (0,-1.15);
    \draw [very thick, ->] (0,-1.75) -- (0,-1.7);
   \draw [very thick, ->](-0.25,-1.5) -- (-0.2,-1.5);
   \draw [very thick, ->](0.25,-1.5) -- (0.2,-1.5);
   % vertex e7
   \draw [very thick, -] (1,-1.5) -- (2,-1.5);
   \draw [very thick, -] (1.5,-2) -- (1.5,-1);
   \Text(56,-33)[]{$\omega_{15}$}; 
   \draw [very thick, ->] (1.5,-1.2) -- (1.5,-1.15);
   \draw [very thick, ->] (1.5,-1.75) -- (1.5,-1.7);
   \draw [very thick, ->](1.2,-1.5) -- (1.15,-1.5);
   \draw [very thick, ->](1.8,-1.5) -- (1.85,-1.5);
% vertex e8
   \draw [very thick, -] (2.5,-1.5) -- (3.5,-1.5);
   \draw [very thick, -] (3,-2) -- (3,-1);
   \Text(99,-33)[]{$\omega_{16}$}; 
   % arrows
   \draw [very thick, ->] (3,-1.25) -- (3,-1.3);
   \draw [very thick, ->] (3,-1.8) -- (3,-1.85);
   \draw [very thick, ->](2.75,-1.5) -- (2.8,-1.5);
   \draw [very thick, ->](3.25,-1.5) -- (3.2,-1.5);
\end{tikzpicture}
\end{center}
\caption{\small The sixteen vertices with their weights $\omega_\alpha$, $\alpha=1, \dots, 16$, 
attached to them. The first six vertices constitute the 
six vertex model with just two-in two-out vertices (the first four with ferro-electric or ferro-magnetic FM order and the 
next two with anti ferroelectric or antiferromagnetic AF order). Adding the next two vertices, with 
four-out and four-in legs, the eight vertex model is built. Finally, the remaining eight vertices 
with three-in and one-out or three-out and one-in arrows drawn in the second row complete the sixteen vertex model.}
\label{fig:uno}
\end{figure}

The {\it six vertex model}~\cite{BaxterBook,Wuphasetransitions,Reshetikhin} was introduced as a model of ferroelectricity. 
It is commonly defined on a square lattice with $N\times N$ vertices.  
Arrows  with two possible orientations are  placed  along  the  links. 
For a lattice with coordination four, there are four edges joining each vertex, see Fig.~\ref{fig:uno}.
The six vertex rule imposes that two arrows point
in  and  two arrows point  out  each vertex.    Depending on the relative orientation of the 
arrows the vertices can have local ferroelectric or anti-ferroelectric order.  Energies, $\epsilon_\alpha$, and,  consequently,  statistical
weights, $\omega_\alpha \propto e^{-\beta \epsilon_\alpha}$ with $\alpha=1,\dots , 6$, are assigned to each vertex.  $\beta=1/(k_BT)$ with $T$ temperature and $k_B$ the 
Boltzmann factor.
Assuming complete 
arrow reversal symmetry only three parameters, $a\equiv \omega_1 =\omega_2$, $b\equiv \omega_3 =\omega_4$, and $c\equiv \omega_5 =\omega_6$, 
are needed to characterise these weights. 
$a$ and $b$ are associated to ferro-electric order and $c$ to anti-ferro-electric order. Under applied fields the 
arrow reversal symmetry is broken and the statistical identity between some of these weights is no longer justified. 
Clearly, as each arrow is shared by two neighbouring vertices correlations can be induced in the 
systems configurations.
The partition function is $Z=\sum_C e^{-\beta\sum_\alpha n_\alpha \epsilon_\alpha}$ where the sum runs over all allowed configurations and $n_\alpha$ is the number of 
vertices of type $\alpha$ in the configuration.

\vspace{0.5cm}

\begin{figure}[h]
\centerline{
\includegraphics[scale=0.5,angle=-90]{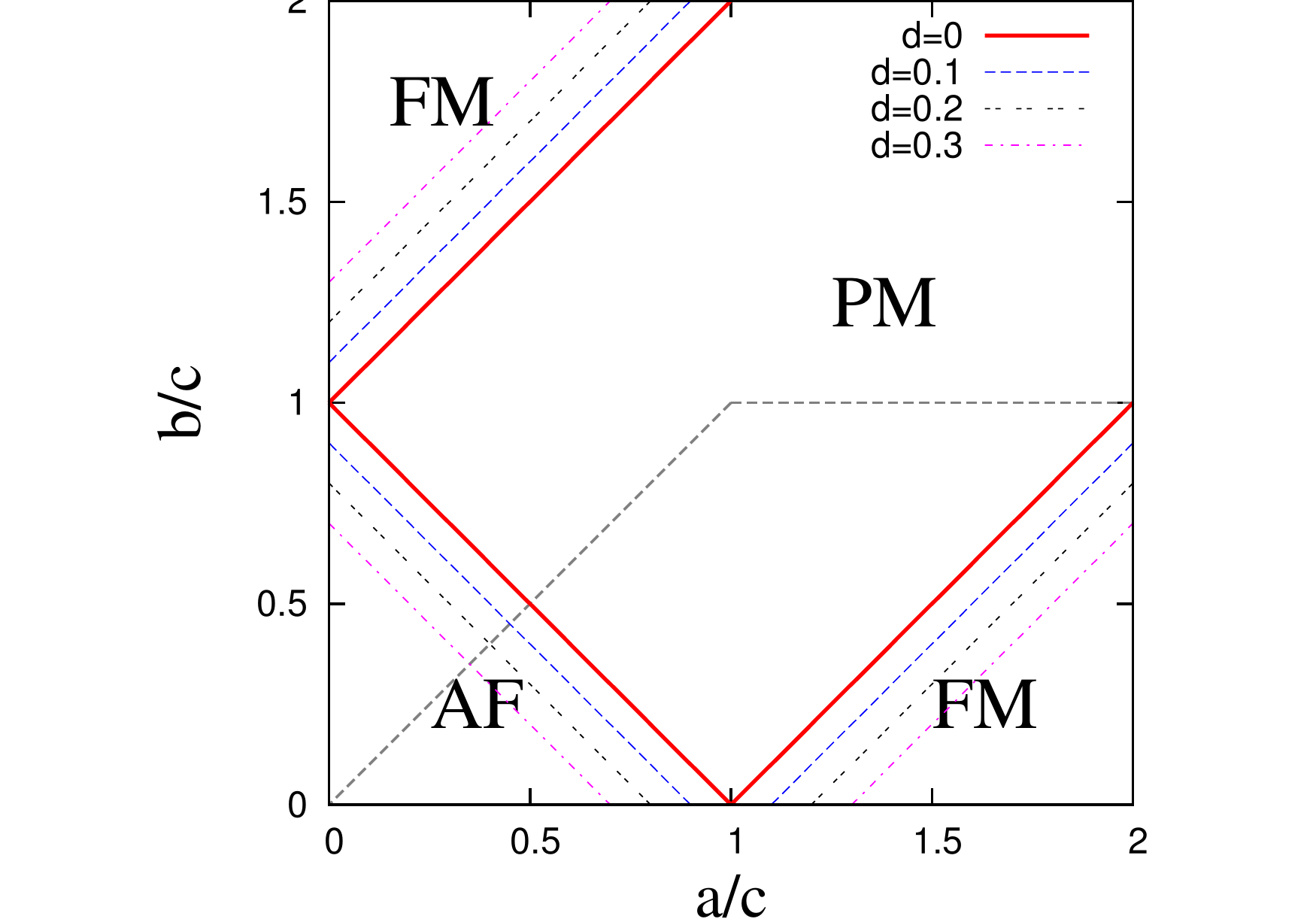}
\hspace{0.5cm}
}
\caption{\small (Colour online.) The phase diagram of the eight-vertex model. The red solid line marks the boundary between different phases in
the six vertex model limit. The dotted inclined (colour) lines are the projections on the $a/c$-$b/c$ plane of the boundaries for various values of $d$. The dashed 
black line shows the parameters for the F and KDP models. Figure taken from~\cite{Foini2013}.}
\label{fig:phase-diagram}
\end{figure}

Lieb solved the six vertex model  
using the {\it transfer matrix} for parameters taking equal values $a=b=c$~\cite{Lieb1967a}, the so-called {\it spin-ice point} in parameter space, 
the choice $a/c=b/c$ or F-model~\cite{Lieb1967b}, and the case $b/c=1$ and $0<a/c<1$ or KDP model~\cite{Lieb1967c}. In particular,  
he computed exactly the macroscopic entropy at the spin-ice point. 
The method was then extended and applied by Sutherland to solve the general case
and the full phase diagram was elucidated, presenting ferroelectric and anti-ferroelectric phases on top of the
disordered (critical) one with power-law decaying correlation functions~\cite{Sutherland1967} (also called a spin-liquid). 
The full phase diagram is shown in Fig.~\ref{fig:phase-diagram}.
The ferroelectric phase is frozen (no fluctuations are permitted) and the 
anti-ferroelectric one is not. The transition lines were found to be of first order between 
disordered and ferroelectric phases and Kosterlitz-Thouless-like between disordered and anti-ferroelectric phases.

\vspace{0.5cm}

The six vertex model is an example of the general field of {\it frustrated magnetism}~\cite{DiepBookCH7short,Balents2010,Gingras2010}. These are classical and quantum systems in
which the interactions in combination with the lattice structure impede the spins to order in an optimal configuration at zero 
temperature. In classical instances, the local minimisation of the interaction energy on a frustrated 
unit gives rise to an exponentially large degeneracy of the ground state and, consequently, a macroscopic residual entropy.
This occurs in the pyrochlore spin-ice Dy$_2$Ti$_2$O$_7$~\cite{Harris1997} in which the spin interactions
are frustrated, similarly to what happens with the proton positions in water ice~\cite{Bernal1933}.  Specific heat measurements using the same sample~\cite{Ramirez1999,Bramwell2001a},
and  the ones performed in water-ice~\cite{Giauque1936}, find an {\it zero-point excess entropy} that is very close to the value that Pauling found with a 
simple counting argument~\cite{Pauling1935}, and even closer to Lieb's exact result for the six vertex model~\cite{Lieb1967a}. Given this magnetic connection, in the rest of this manuscript I will use the magnetic terminology with the two ordered phases being called ferromagnetic and anti-ferromagnetic.

Boundary conditions do not usually affect the bulk behaviour of  macroscopic samples.
Frustrated models can provide exceptions to this rule and the six vertex model is indeed one 
such example. Special border rules used in the six vertex case are the so-called 
{\it domain-wall boundary conditions} in which all arrows on the bottom and top boundaries
enter the lattice while  all  arrows  on  the  left  and  right  boundaries  exit the lattice.
The partition function of the six vertex model under these conditions satisfies a recurrence relation that leads to
a determinant formula used to derive the free-energy densities in all phases~\cite{Izergin87,Kuperberg96,KorepinZinnJustin00}.
Interestingly enough, although the phase diagram remains unchanged, the order of the disordered-ferroelectric 
transition becomes continuous. Moreover, the  free-energy  densities  in  the
disordered and antiferromagnetic phases, are different from the ones for periodic
boundary  conditions,  even  in  the  thermodynamic  limit.   This difference is intimately linked to a macroscopic phase separation 
in real space induced by the boundary conditions. For example, for bulk parameters in the disordered phase 
an {\it arctic curve} separates an external frozen domain from an internal temperate one, both with finite spatial density.
Such an arctic curve  first appeared in the study of domino
tilings of Aztec diamonds~\cite{Elkies1992a,Elkies1992b,Cohn1996,Jockusch1998},  
then in lozenge tilings of large hexagons~\cite{Cohn1998, Borodin2010},
and later in more general dimer~\cite{Kenyon} and vertex models~\cite{Cohn1996,Jockusch1998, ColomoPronko-proc, ColomoPronkoZinnJustin2010, ColomoPronko2010a, Colomoetal2011}.  In these systems phase separation exists for a wide choice of  fixed boundary conditions and parameter values in the model definition.

\vspace{0.25cm}
\noindent
{\bf The eight vertex model}
\vspace{0.25cm}

The strict two-in two-out condition can be partially lifted to allow for vertices with four-in or four-out 
arrows and thus define the {\it eight vertex model}, with vertices drawn in the first line in Fig.~\ref{fig:uno}.
This case is less constrained but also solvable analytically.  The phase diagram 
still has ordered and disordered phases although the latter is no longer critical and the 
transition lines towards the ferromagnetically ordered phases are now continuous~\cite{BaxterBook} (see Fig.~\ref{fig:phase-diagram}
where the projection on the $a/c$-$b/c$ plane is shown for different values of the parameter $d = \omega_7 =  \omega_8$).
The peculiarity of this problem is that the critical exponents are continuous functions 
of a particular combination of the vertex weights $\omega_\alpha$. 
At first, this fact seemed to contradict the {\it universality 
hypothesis}. Kadanoff and Wegner introduced a mapping to a spin-model
with multi-spin interactions that shed light on the apparent violation of universality~\cite{Kadanoff1971}. I will briefly 
explain it below. 

The six and eight vertex  models admit a large number of mappings to other also very interesting statistical and quantum physical systems:
three-coloring problems, random tilings, interacting dimer coverings, surface growth, alternating sign matrices and 
quantum spin chains (with the equilibrium properties of the 
six vertex model being equivalent to the ones of the XXZ spin chain, and the ones of the eight vertex model corresponding to the ones of the 
XYZ spin chain)~\cite{BaxterBook}. The Coulomb gas method and conformal-field theory techniques have added significant insight 
into the phase transition and critical properties of these systems. 

\vspace{0.25cm}
\noindent
{\bf The sixteen vertex model}
\vspace{0.25cm}

The {\it sixteen vertex model} treats on an equal footing, although with different probability weights $\omega_\alpha$, all possible four leg vertices on a square
lattice. All these vertices are depicted in Fig.~\ref{fig:uno}.  Under no external applied field the model is assumed to be spin-reversal 
symmetric and only eight such parameters exist, $\omega_{2k+1}=\omega_{2k}$ with $k=0, \dots, 7$. 
On the contrary, when fields are applied 
the degeneracy between certain energies is lifted and the probability weights can differ.

Quite naturally, much less is known about the equilibrium  properties of a
generic vertex model that breaks integrability. Indeed, as soon as the integrability conditions  are lifted,  
the exact techniques  are no longer useful and the mappings to other solvable problems also break down.

\vspace{0.25cm}
\noindent
{\bf Mapping to a classical spin model with multi-spin interactions}
\vspace{0.25cm}

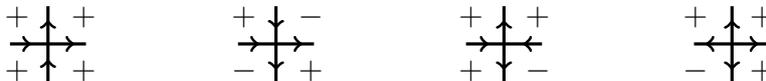
\begin{figure}[h]
\begin{center}
\begin{tikzpicture}
% vertex a1 
   \draw [very thick, -] (-8,0) -- (-7,0);
   \draw [very thick, -] (-7.5,-0.5) -- (-7.5,0.5);
   \Text(-200,10)[]{$+$}; 
   \Text(-225,10)[]{$+$}; 
   \Text(-200,-10)[]{$+$}; 
   \Text(-225,-10)[]{$+$}; 
   % arrows
   \draw [very thick, ->](-7.2,0) -- (-7.15,0);
   \draw [very thick, ->](-7.75,0) -- (-7.72,0);
   \draw [very thick, ->] (-7.5,-0.25) -- (-7.5,-0.2);
   \draw [very thick, ->] (-7.5,0.27) -- (-7.5,0.32);
   % vertex b1
   \draw [very thick, -] (-5,0) -- (-4,0);
   \draw [very thick, -] (-4.5,-0.5) -- (-4.5,0.5);
   \Text(-115,10)[]{$-$}; 
   \Text(-140,10)[]{$+$}; 
   \Text(-115,-10)[]{$+$}; 
   \Text(-140,-10)[]{$-$}; 
   % arrows
   \draw [very thick, ->](-4.2,0) -- (-4.15,0);
   \draw [very thick, ->](-4.75,0) -- (-4.72,0);
   \draw [very thick, ->] (-4.5,-0.3) -- (-4.5,-0.35);
   \draw [very thick, ->] (-4.5,0.25) -- (-4.5,0.2);
   % vertex c1
   \draw [very thick, -] (-2,0) -- (-1,0);
   \draw [very thick, -] (-1.5,-0.5) -- (-1.5,0.5);
   %arrows
   \draw [very thick, ->] (-1.5,-0.3) -- (-1.5,-0.35);
   \draw [very thick, ->] (-1.5,0.27) -- (-1.5,0.32);
   \draw [very thick, ->](-1.75,0) -- (-1.7,0);
   \draw [very thick, ->](-1.2,0) -- (-1.25,0);
   \Text(-30,10)[]{$+$}; 
   \Text(-55,10)[]{$+$}; 
   \Text(-30,-10)[]{$-$}; 
   \Text(-55,-10)[]{$+$}; 
   % vertex d1
   \draw [very thick, -] (1,0) -- (2,0);
   \draw [very thick, -] (1.5,-0.5) -- (1.5,0.5);
   %arrows
   \draw [very thick, ->] (1.5,-0.3) -- (1.5,-0.35);
   \draw [very thick, ->] (1.5,0.27) -- (1.5,0.32);
   \draw [very thick, ->](1.2,0) -- (1.15,0);
   \draw [very thick, ->](1.8,0) -- (1.85,0);
    \Text(54,10)[]{$+$}; 
    \Text(29,10)[]{$+$}; 
    \Text(54,-10)[]{$+$}; 
    \Text(29,-10)[]{$-$};   
\end{tikzpicture}
\end{center}
\caption{\small The mapping between vertex and spin configurations on the dual lattice~\cite{Kadanoff1971}. Only four vertices out of the 
eight vertices in the eight vertex model are shown and only one spin configuration for each vertex is drawn. The other possibilities are straightforward.}
\label{fig:dos}
\end{figure}

Kadanoff and Wegner~\cite{Kadanoff1971}, and simultaneously Wu~\cite{Wu1971}, 
showed that the eight-vertex model on a square lattice 
is equivalent to a classical  Ising model on its (also square) dual lattice.  The equivalence goes as follows.
First, we note that there are 
eight different vertices in the eight vertex model. With four spins, located at the
centres of the adjacent plaquettes to a vertex that are sites of the dual lattice, one has $2^4=16$ different configurations.
There will then be a degeneracy in the mapping, such that two spin configurations will correspond to one vertex 
configuration. The criterium for the mapping is indicated in Fig.~\ref{fig:dos} on four examples: \\
-- an arrow pointing up (down) on a vertical link is equivalent to two parallel (antiparallel) spins located at the centre of the 
adjacent plaquettes, that is to say, on the closest sites on the dual lattice.\\
-- an arrow pointing right (left) on a horizontal edge is equivalent to two parallel (antiparallel) spins located at interstitial sites 
of the lattice.

\vspace{0.5cm}

The more general eight vertex model  has eight independent parameters, 
$\omega_1$, $\omega_2$, $\omega_3$, $\omega_4$, $\omega_5$, $\omega_6$, $\omega_7$, $\omega_8$.
Under periodic boundary conditions the global number of outgoing and incoming arrows must be the same.
Indeed, there cannot be sources or sinks of arrows and, as the antiferromagnetic ($c$) and four-in or four-out vertices do 
act as local sources or sinks, the conditions $\omega_5=\omega_6$ and $\omega_7=\omega_8$ must hold. The simplest 
Hamiltonian with local interactions on a square plaquette and six parameters is
\begin{eqnarray}
H_{8v}(\{s_i\}) \!\! &=& \!\! - J_0 N - \sum_{ij} 
\Big{(} 
J_n^x \sigma_{i,j} \sigma_{i,j+1} + J_{n}^y \sigma_{i,j} \sigma_{i+1,j}
+ J_{nn} \sigma_{i,j+1} \sigma_{i+1,j} 
\nonumber\\ 
&&
\qquad\qquad\quad
+ J'_{nn} \sigma_{i,j} \sigma_{i+1,j+1} + J_4 \sigma_{i,j} \sigma_{i,j+1} \sigma_{i+1,j} \sigma_{i+1,j+1}
\Big{)}
\; . 
\label{eq:hamiltonian-kadanoff}
\end{eqnarray}
This model has anisotropic nearest-neighbour interactions mediated by $J_n^x$ and $J_n^y$;
diagonal, next-nearest-neighbour interactions, with coupling strengths $J_{nn}$ and $J_{nn}'$ depending on the 
direction of the diagonal; and plaquette four-spin interactions with exchange $J_4$. The relation between the vertex weights 
$\omega_\alpha$ and the coupling constants $J$s are given by the evaluation of 
$\omega_\alpha = e^{-\beta H_{8v}(\{s_i\})}$ for the spin configurations corresponding to each vertex
(the normalisation $Z$ can be absorbed in the parameter $J_0$).

In the particular case $\omega_1=\omega_2=a$, $\omega_3=\omega_4=b$
the first-neighbour couplings vanish, $J_n^x=J_n^y=0$. The model is then a pair of Ising models
on two square lattices coupled by the four-spin interaction.
The critical exponents are parametrized by $\tan(\mu/2) = [cd/(ab)]^{1/2} = e^{-2J_4}$
and it is clear that the four body interaction is responsible for their parameter dependence.
For example, the specific heat behaves as~\cite{Kadanoff1971} 
\begin{equation}
C_v \simeq \epsilon^{-\alpha}
\; , 
\qquad\quad
\epsilon \sim (b+c+d-a)/a 
\; ,
\qquad\quad
\sin\frac{\pi\alpha}{4(1-\frac12\alpha)} = \tanh 2J_4
\end{equation}
close to one of the transition lines.

The mapping can be taken one step further and be extended to 
the sixteen vertex model with $a$, $b$, $c$, $d$ and equal weight for all three-in one-out and three-out one-in vertices 
parametrized by $e$~\cite{Wuphasetransitions}. Place now an Ising variable on the middle point of each edge between two 
vertices. In this way, there are as many spins as links on the original square lattice. Each vertex has an 
up, $\sigma_u$, a down, $\sigma_d$, a right, $\sigma_r$, and a left, $\sigma_l$, spin attached to it.
If each of these spins interacts with  its nearest-neighbour, its next-nearest neighbour and over the plaquette
that they form, the energy of the signalled vertex is
\begin{eqnarray}
H^{(v)}_{16v}(\{s_i\}) \!\! &=& \!\! - J_0 - 
J_n^x \left( \sigma_{l} \sigma_{u} + \sigma_d \sigma_r \right) 
-
J_{n}^y \left( \sigma_{l} \sigma_{d} + \sigma_u \sigma_r  \right)
\nonumber\\ 
&&
\qquad\qquad\quad
- J_{nn} \sigma_{u} \sigma_{l} 
- J'_{nn} \sigma_l \sigma_r + 
J_4 \sigma_{u} \sigma_{d} \sigma_{l} \sigma_{r}
\; . 
\label{eq:sixteen-vertex}
\end{eqnarray}
There is a special relation between the parameters, $e^4 = abcd$, such that $J_4=0$ and only two-body interactions 
remain. Some exact results for the equilibrium of this case are known.

Ising spin models with plaquette interactions acquired an interest {\it per se} after the work of Kadanoff and Wegner, and Wu, and many papers
were devoted to the study of their phase diagram and critical properties with different techniques, including finite-size scaling~\cite{Barber79}, perturbation theory, low- and high- temperature expansions, 
field theoretical tools~\cite{Minami93}, and Monte Carlo simulations~\cite{Landau80,BinderLandau80,LandauBinder85}.
The cluster variation method was also used to study this problem~\cite{BuzanoPretti,Cirillo99}. Plaquette spin models were used to mimic glassy behaviour
within the description provided by kinetically constrained models,  
especially by Jack and co-workers~\cite{Jack05a,Jack05b,Jack06,Turner15,Jack16}.
  
\section{Artificial spin-ice}

Two-dimensional Ising-like ice models found a nice experimental counterpart recently when it became 
possible to manufacture artificial samples with arrays of single-domain ferromagnetic nano-islands 
frustrated by dipolar interactions~\cite{Heyderman2013,NisoliMoessnerSchiffer2013}. 
In their simplest setting {\it artificial spin-ice} (ASI) are 2D arrays of elongated single-domain 
permalloy islands whose shape anisotropy defines Ising-like spins arranged along the edges of a regular square lattice. 
Other lattice geometries can be drawn in the laboratory as well.
Spins interact through dipolar exchanges and the dominant contributions are the ones between neighbouring islands across a given vertex. No configuration of the surrounding spins can minimize all pairwise dipole-dipole interactions
on a vertex. The interaction parameters can be 
precisely engineered -- by tuning the distance between islands, i.e., the lattice constant, the height between layers or by applying external fields.
In this way one can select the phase into which the system should settle in~\cite{Wang2006,Nisoli07,Morgan10}.
One of the main goals of the research on artificial spin-ice is to develop new
materials that could improve the performance of data storage and data processing devices.

 In samples with no height offset, the 2D square symmetry defines five relevant vertex types 
of increasing energy, where the c vertices take the lowest value, leading to a ground state with staggered 
c-AFM order~\cite{Morgan13}.
However, the relative energies of the different vertex configurations could be tuned differently 
in such a way that the ground state displayed other types of order or be even disordered.

In the experiments in~\cite{Morgan10,Morgan13} the thickness of the magnetic islands grows by deposition
 (while temperature and all other external parameters are kept constant within experimental accuracy). 
 The Ising spins flip by thermal fluctuations during the growth process. As the energy barrier for single 
 spin flips increases with the size of the islands, once a certain thickness is reached the barrier crosses-over the 
energy provided by the bath,  
$k_BT$, and the spins freeze (experiments are usually performed at room temperature). 
At the end of the growth process, the frozen spin configurations are imaged with 
magnetic force microscopy~\cite{Wang2006}, or other techniques~\cite{PhysRevB.83.174431,Remhof08,PhysRevB.78.144402,Ladak10,Li10a,Li10b,PhysRevLett.111.057204}
and the number of vertices of each kind are counted. 
A statistical analysis of the microscopic configurations is carried out and
averaged values  (with statistical errors) are evaluated. The configurations thus sampled are not necessarily the ones 
of thermal Boltzmann equilibrium at the working temperature and several groups have tried to find an effective statistical 
measure to describe them~\cite{Morgan13,Nisoli10}.

One can model 2D ASI by taking into account dipolar interactions~\cite{Harris1997,Bramwell2001a,Moller2006,Wysin12} or by using a simpler vertex model.
If the latter choice is made, the complete vertex model on a 
square lattice, where all kinds of vertices are allowed, should be used. The latter route was the one that I followed in recent years, 
and I summarise some of the results that we found in the next Section.

\section{Results}

In this Section we present some recent  results on the use of vertex models to 
describe bidimensional spin-ice samples.

\vspace{0.25cm}
%\newpage
\noindent
{\bf Equilibrium properties of the sixteen vertex model}
\vspace{0.25cm}

Approximate methods, such as the Bethe-Peierls  approximation~\cite{Bethe35} 
and its modern versions, like the cavity method
and the belief propagation algorithm~\cite{Pearl,Yedidia2003,CVMreview,MezardMontanari}, 
turned out to be of great help to obtain the equilibrium properties of generic vertex models~\cite{Levis2013a,Foini2013,CGP1996a,CGP2012}.  
In~\cite{Levis2013a,Foini2013} we introduced a suitable Bethe-Peierls approximation, defined on a well-chosen tree of plaquettes, 
and we derived self-consistent equations on such a tree, the fixed points of which yield 
the exact solution of the model in this approximation. Surprisingly enough,  the method gave
very  accurate,  sometimes  even  exact,  results  when  applied  to  the  integrable  six  and
eight  vertex  cases.  For instance, the location in parameter space of the transition lines is captured exactly in the six and eight vertex
models. The first order character of the transition between disordered and ferromagnetic phases in the six vertex model 
is also found. However, the disordered phase, named PM for paramagnetic in Fig.~\ref{fig:phase-diagram}, is not critical but just a conventional 
high temperature phase. Consequently, 
the Kosterlitz-Thouless transition between disordered and ferromagnetic phases in the same model 
is mistaken by a second order one. The projection of the phase diagram for the eight vertex model on the plane $a/c$-$b/c$ is 
shown in Fig.~\ref{fig:phase-diagram} that is extracted from Ref.~\cite{Foini2013}.

For the sixteen vertex model the method allowed us to describe all expected phases and to 
unveil some of their properties, such as the presence of anisotropic equilibrium fluctuations in the symmetry broken phases.
The predictions of the Bethe-Peierls approximation were 
confronted to  Monte Carlo (MC) simulations of the finite-dimensional system with very good agreement. 

For small values of the probability weight of the defects, that is to say, $d \ll e\ll a, \, b, \, c$, the ordered anti-ferromagnetic and 
ferromagnetic phases survive as well as the disordered phase. The latter loses its critical properties and the 
ferromagnetic phase is no longer frozen. This is the parameter regime that is relevant for most experiments performed
with artificial spin-ice samples, as we explain below.

\vspace{0.25cm}
\noindent
{\bf Artificial spin-ice}
\vspace{0.25cm}

In~\cite{Levis2013a} we made contact with experiments by choosing  parameters in  the  sixteen  vertex  model close  to
the ones of artificial spin-ice samples~\cite{Heyderman2013,NisoliMoessnerSchiffer2013}
obtained by gradual 
deposition of magnetic material on square patterns with different lattice constant and varying under-layer disorder.
The single vertex energies can be estimated to be 
$\epsilon_c=(-2\sqrt{2}+1)/\ell$, 
$\epsilon_a=\epsilon_b=-1/\ell$,
$\epsilon_e=0$,
$\epsilon_d=(4\sqrt{2}+2)/\ell$  with 
$\ell$ the length of the individual magnets (edges on the lattice). These expressions were obtained by Nisoli {\it et al.}
modelling the arrows with  two opposite charges and taking into account the electrostatic energy between them~\cite{Nisoli10}.

\begin{figure}[t]
\centerline{
\includegraphics[scale=0.9]{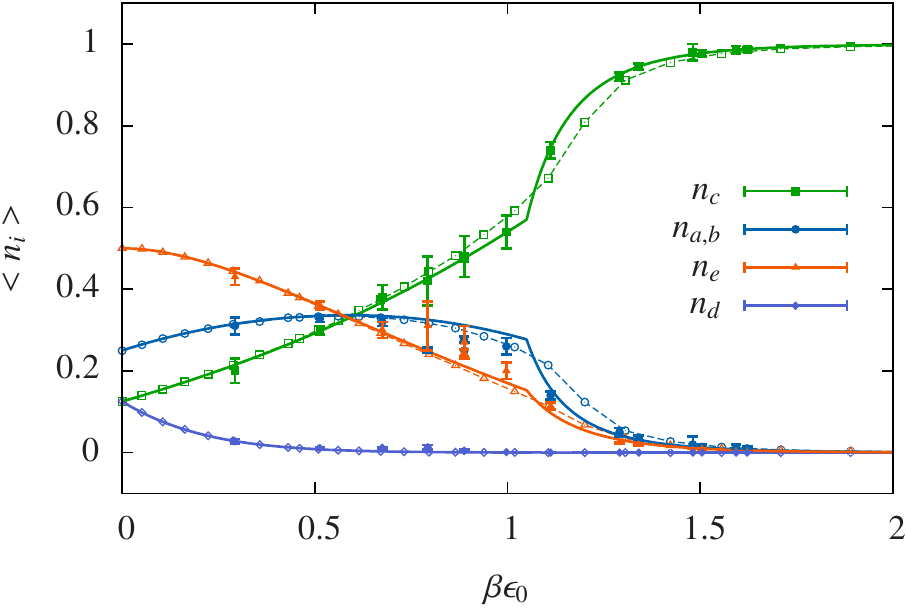}
\hspace{0.5cm}
}
\vspace{0.4cm}
\centerline{
\includegraphics[scale=0.17]{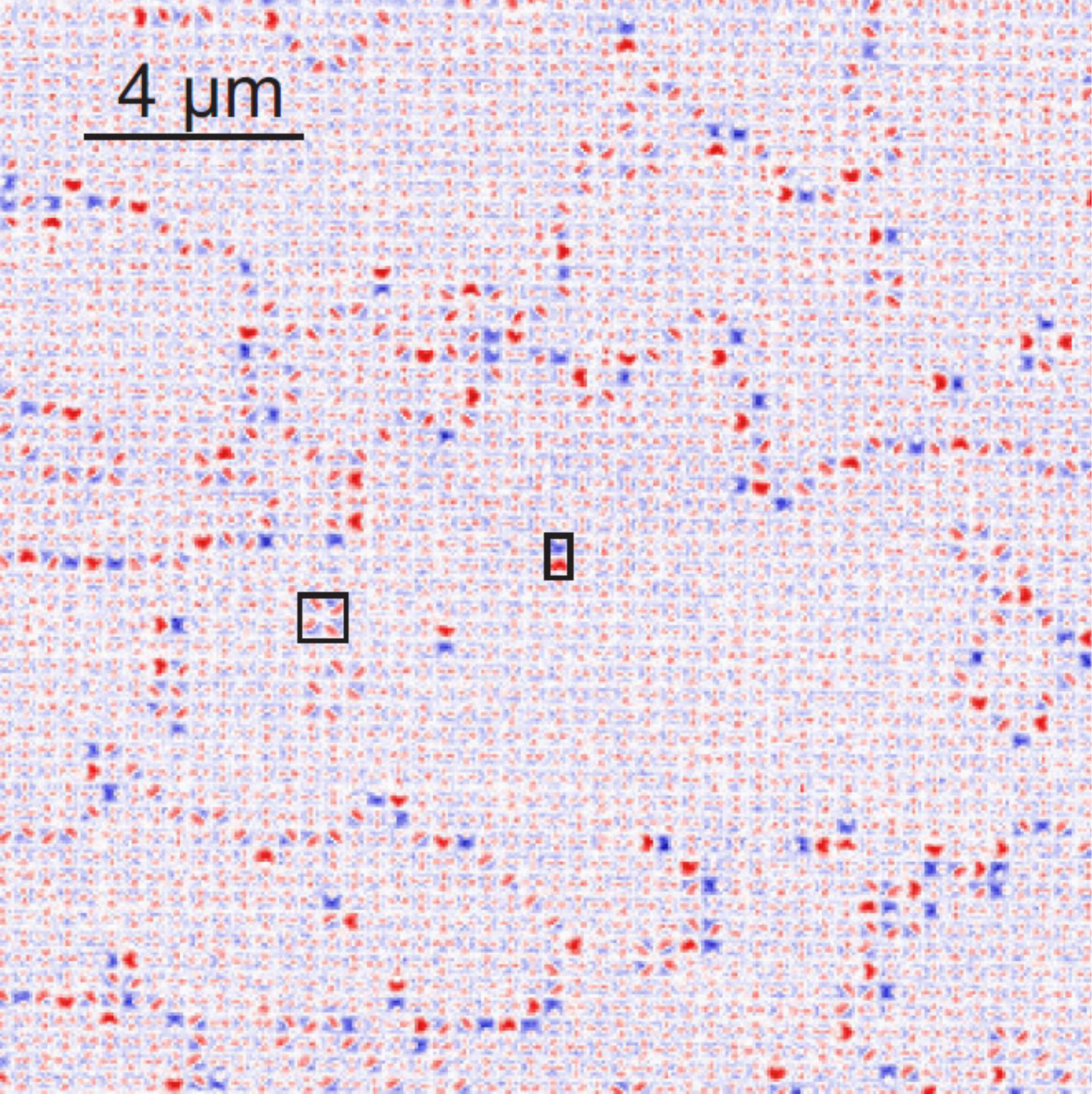}
\includegraphics[scale=0.27]{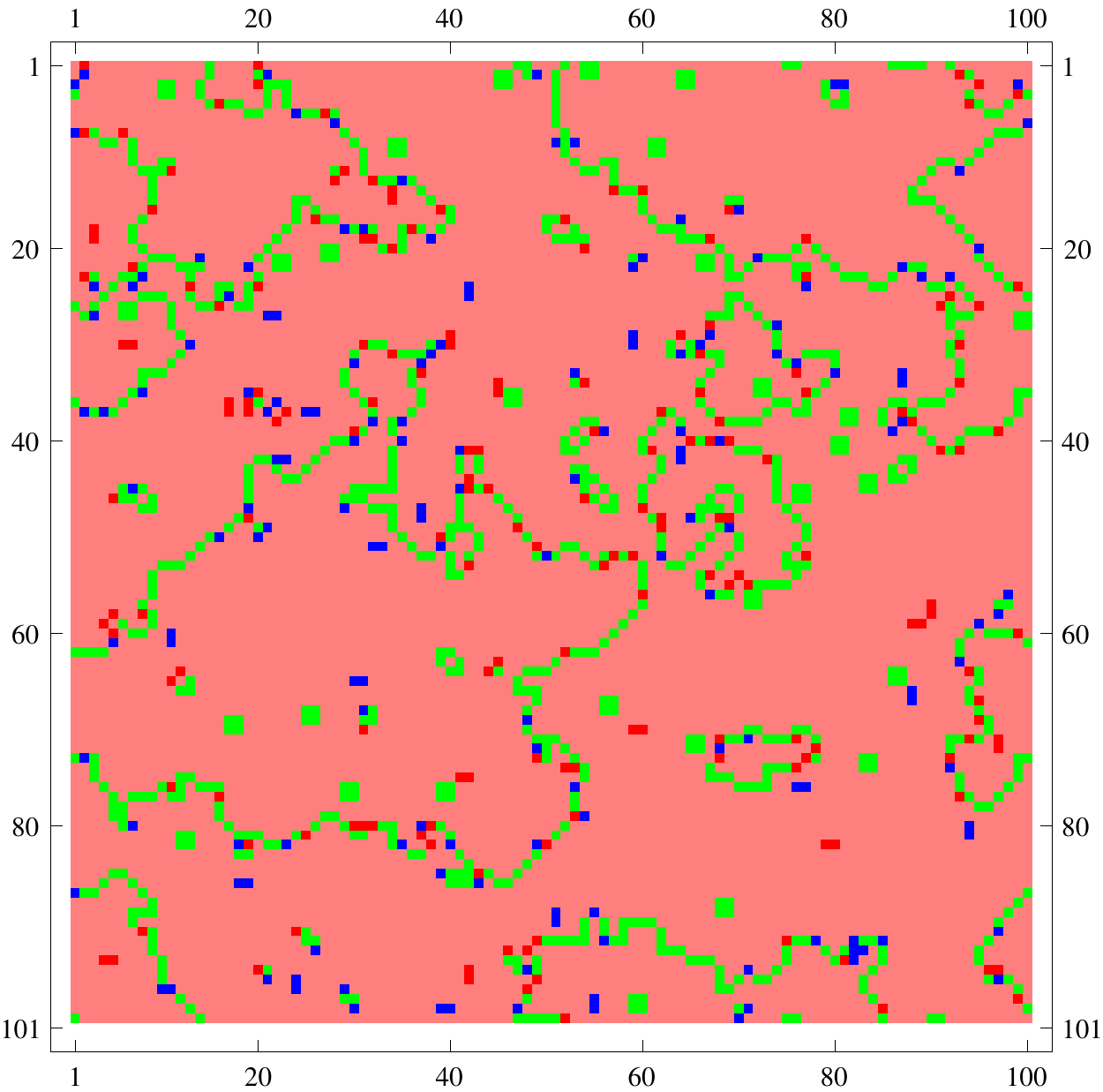}
\includegraphics[scale=0.27]{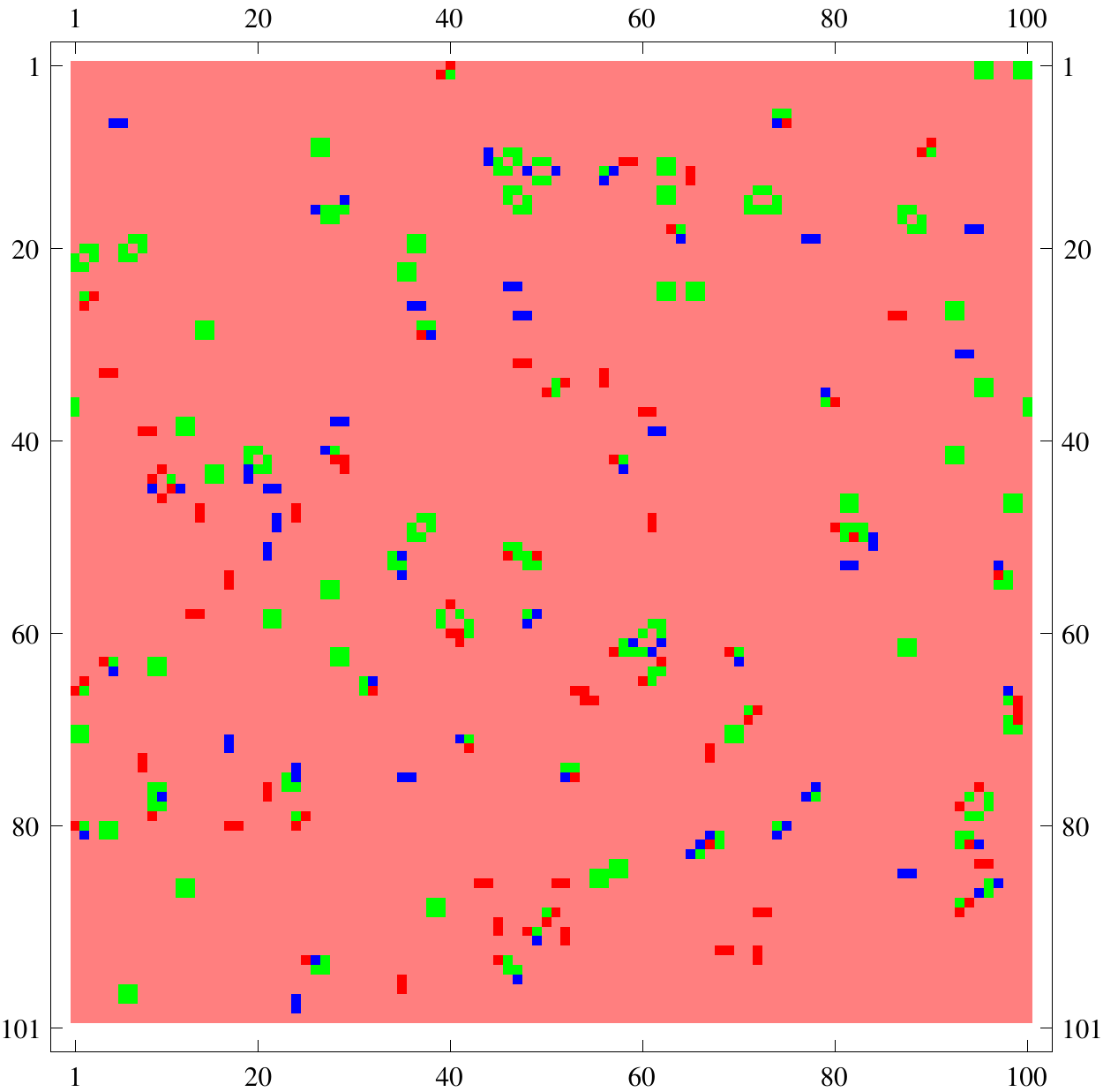}
}
\vspace{0.25cm}
\caption{\small (Colour online.)
Upper panel: figure extracted from~\cite{Levis2013a}.
The average densities of vertices of different type
as a function of $\beta \epsilon_0$ with $\beta$ the inverse temperature and $\epsilon_0$ a reference vertex energy,
see the text. 
Full symbols with error bars are experimental
data~\cite{Morgan10}. Empty symbols with dotted lines correspond to
the equilibrium CTMC (Continuous Time Monte Carlo) data. 
The cluster variational Bethe-Peierls analytic solution of the sixteen-vertex model is shown with solid lines.
In the lower panels we show a typical experimental configuration, figure taken from~\cite{Budrikis12b} ({\copyright \;\; IOP Publishing \& Deutsche Physikalische Gesellschaft, CC BY-NC-SA}), and 
two numerical configurations taken from~\cite{Levis2013a}. The snapshot in the central panel the first one is out of equilibrium and the one in the 
third panel is 
in equilibrium. All these snapshots are for parameters in the anti-ferromagnetic phase.
}
\label{fig:vertex-densities}
\end{figure}
The local energies are then ordered as $\epsilon_c< \epsilon_{a,b} < \epsilon_e< \epsilon_d$, leading to $d<e<a,b<c$.
For these parameters, 
the six vertex model predicts a second-order phase transition from a conventional high temperature (or large lattice constant, strong disorder) 
disordered phase to a low temperature (or small lattice constant, weak disorder) staggered antiferromagnetic phase that was not 
taken into account in previous analysis of the experimental data.  In the upper panel in Fig.~\ref{fig:vertex-densities} we should the 
dependence of the vertex densities as a function of inverse temperature times 
a reference vertex energy, $\epsilon_0\equiv \epsilon_c$. Temperature is fixed in the experiments but samples with different 
$\ell$ or substrate are studied changing therefore the reference (and other) vertex energy parameters.
We include in the figure the experimental data (with full symbols), equilibrium Monte Carlo data (with open symbols joined with 
dotted lines), and the analytic solution of the sixteen vertex model (with solid lines). The agreement between
experimental data and the model results is very good away 
from the critical point, implying that the experimental samples of~\cite{Morgan10,Morgan13} are at -- or at least very close to -- thermal equilibrium
for such parameters. However, deviations are seen close to criticality were, most probably, the samples have not had enough time
to equilibrate during preparation. This interpretation does not require a fitting parameter, such as the effective temperature introduced in~\cite{Nisoli10,Morgan13}. The lower panels in the same figure show experimental and numerical snapshots.
The left picture is taken from Ref.~\cite{Morgan13}.
The homogenous-looking part corresponds to antiferromagnetic order and the domain-walls and lines of defects are 
darker. The other two pictures are taken from Ref.~\cite{Levis2013a}. The uniform looking regions are anti-ferromagnetically 
ordered and the defects and vertices of other kind are shown with different colours. The central picture is out of equilibrium. The 
number of defects is larger and they are mostly organised in domain walls. 
The spatial arrangement of vertices in near-critical artificial spin-ice should be studied in more detail and confronted to the correlations 
expected in equilibrium or after quenches to understand how far form equilibrium the samples are.

\vspace{0.25cm}
\noindent
{\bf Domain wall boundary conditions}

\vspace{0.25cm}

The effect of fixed boundary conditions of the domain-wall type were studied in Ref.~\cite{CuGoPe} with the Bethe-Peierls or  cluster variational 
method. Interestingly enough, this method allows one to obtain the arctic curves with the same degree of
difficulty for all values of the parameters in the model, be them in the disordered or in the antiferromagnetic phase. 
Moreover, the method can be adapted to deal with lattices with rectangular shape.
The curves found with this admittedly only approximate method are remarkably close to the exact ones when 
these are known. It is quite surprising indeed that a `mean-field' method can capture real-space phase separation
with such a degree of precision.

\vspace{0.5cm}

\begin{figure}[h]
\centerline{
\includegraphics[scale=0.205]{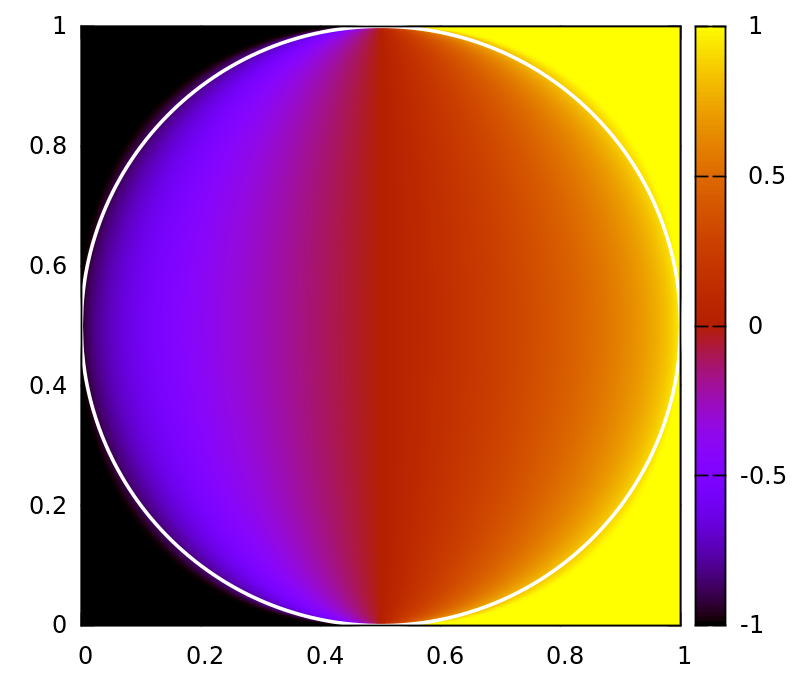}
}
\caption{\small (Colour online.)
Polarization of the horizontal edges at the
so-called free-fermion case, $\Delta \equiv (a^2+b^2-c^2)/(2ab) = 0$, with $a
  = b$, on a square lattice with $1024 \times 1024$ lattice vertices, computed with the Bethe-Peierls approximation. 
  The white line is the exact arctic circle. This figure is taken from Ref.~\cite{CuGoPe}.}
\end{figure}

\vspace{0.25cm}
\noindent
{\bf Order by disorder}

\vspace{0.25cm}

{\it Order-by-disorder} (ObD) is the mechanism whereby a system with a non-trivially 
degenerate ground state develops long-range order by 
the effect of classical or quantum fluctuations~\cite{Chalker}. 
More precisely, a huge disproportion in the density of low-energy 
excitations associated with particular ground states that are ordered suffices to select them as soon as an 
infinitesimal temperature is switched on. 
From a theoretical point of view, the ObD mechanism was first exhibited in the 
classical fully frustrated domino model~\cite{Villain} but it 
is a relatively common occurrence in 
geometrically frustrated spin models. However, there is still no definitive experimental evidence for it. 
The difficulty lies in establishing whether the selection of order is due to the 
ObD mechanism, or whether the reason for ordering 
is the contribution of terms  not taken into account in the Hamiltonian model
that actually lift the ground state degeneracy.
In a recent Letter we argued that it should be possible to observe thermal ObD in 2D spin-ice samples with 
parameters such that 
the preferred anti-ferromagnetic staggered order is inhibited by a magnetic field~\cite{Guruciaga16}. 

\vspace{0.5cm}

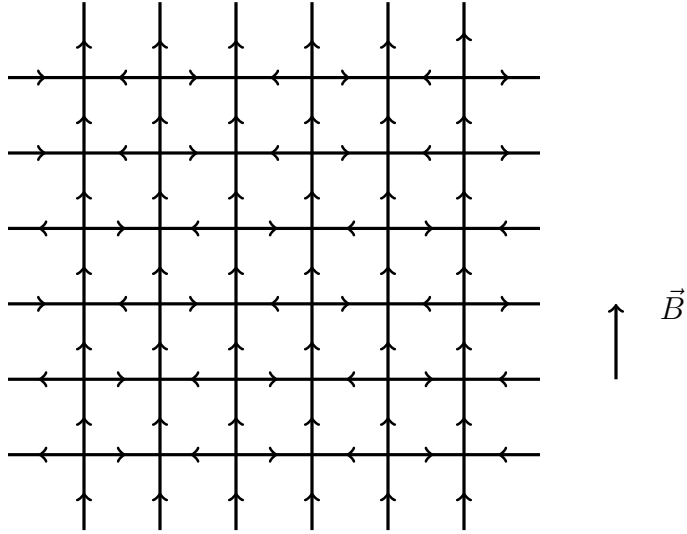
\begin{figure}[h]
\begin{center}
\begin{tikzpicture}
% vertical lines 
   \draw [very thick, -] (-3,-3) -- (-3,4);
% arrows
         \draw [very thick, ->] (-3,-2.55) -- (-3,-2.5);
         \draw [very thick, ->] (-3,-1.55) -- (-3,-1.5);
         \draw [very thick, ->] (-3,-0.55) -- (-3,-0.5);
         \draw [very thick, ->] (-3, 0.45) -- (-3,0.5);
         \draw [very thick, ->] (-3,1.45) -- (-3,1.5);
         \draw [very thick, ->] (-3,2.45) -- (-3,2.5);
         \draw [very thick, ->] (-3,3.45) -- (-3,3.5);
% vertical lines 
    \draw [very thick, -] (-2,-3) -- (-2,4);
% arrows
         \draw [very thick, ->] (-2,-2.55) -- (-2,-2.5);
         \draw [very thick, ->] (-2,-1.55) -- (-2,-1.5);
         \draw [very thick, ->] (-2,-0.55) -- (-2,-0.5);
         \draw [very thick, ->] (-2, 0.45) -- (-2,0.5);
         \draw [very thick, ->] (-2,1.45) -- (-2,1.5);
         \draw [very thick, ->] (-2,2.45) -- (-2,2.5);
         \draw [very thick, ->] (-2,3.45) -- (-2,3.5);
% vertical line
    \draw [very thick, -] (-1,-3) -- (-1,4);
% arrows
        \draw [very thick, ->] (-1,-2.55) -- (-1,-2.5);
         \draw [very thick, ->] (-1,-1.55) -- (-1,-1.5);
         \draw [very thick, ->] (-1,-0.55) -- (-1,-0.5);
         \draw [very thick, ->] (-1, 0.45) -- (-1,0.5);
         \draw [very thick, ->] (-1,1.45) -- (-1,1.5);
         \draw [very thick, ->] (-1,2.45) -- (-1,2.5);
         \draw [very thick, ->] (-1,3.45) -- (-1,3.5);
% vertical line
    \draw [very thick, -] (0,-3) -- (0,4);
% arrows
        \draw [very thick, ->] (0,-2.55) -- (0,-2.5);
         \draw [very thick, ->] (0,-1.55) -- (0,-1.5);
         \draw [very thick, ->] (0,-0.55) -- (0,-0.5);
         \draw [very thick, ->] (0, 0.45) -- (0,0.5);
         \draw [very thick, ->] (0,1.45) -- (0,1.5);
         \draw [very thick, ->] (0,2.45) -- (0,2.5);
         \draw [very thick, ->] (0,3.45) -- (0,3.5);
% vertical line
    \draw [very thick, -] (1,-3) -- (1,4);
% arrows
        \draw [very thick, ->] (1,-2.55) -- (1,-2.5);
         \draw [very thick, ->] (1,-1.55) -- (1,-1.5);
         \draw [very thick, ->] (1,-0.55) -- (1,-0.5);
         \draw [very thick, ->] (1, 0.45) -- (1,0.5);
         \draw [very thick, ->] (1,1.45) -- (1,1.5);
         \draw [very thick, ->] (1,2.45) -- (1,2.5);
         \draw [very thick, ->] (1,3.45) -- (1,3.5);
% vertical line
    \draw [very thick, -] (2,-3) -- (2,4);
% arrows
        \draw [very thick, ->] (2,-2.55) -- (2,-2.5);
         \draw [very thick, ->] (2,-1.55) -- (2,-1.5);
         \draw [very thick, ->] (2,-0.55) -- (2,-0.5);
         \draw [very thick, ->] (2, 0.45) -- (2,0.5);
         \draw [very thick, ->] (2,1.45) -- (2,1.5);
         \draw [very thick, ->] (2,2.45) -- (2,2.5);
         \draw [very thick, ->] (2,3.55) -- (2,3.6);
  % horizontal lines              
   \draw [very thick, -] (-4,3) -- (3,3); 
   % arrows
        \draw [very thick, ->] (-3.55,3) -- (-3.5,3);
         \draw [very thick, ->] (-2.5,3) -- (-2.55,3);
         \draw [very thick, ->] (-1.55,3) -- (-1.5,3);
         \draw [very thick, ->] (-0.5,3) -- (-0.55,3);
         \draw [very thick, ->] (0.45,3) -- (0.5,3);
         \draw [very thick, ->] (1.5,3) -- (1.45,3);
         \draw [very thick, ->] (2.55,3) -- (2.6,3);
% horizontal lines      
   \draw [very thick, -] (-4,2) -- (3,2);
     % arrows
        \draw [very thick, ->] (-3.55,2) -- (-3.5,2);
         \draw [very thick, ->] (-2.5,2) -- (-2.55,2);
         \draw [very thick, ->] (-1.55,2) -- (-1.5,2);
         \draw [very thick, ->] (-0.5,2) -- (-0.55,2);
         \draw [very thick, ->] (0.45,2) -- (0.5,2);
         \draw [very thick, ->] (1.5,2) -- (1.45,2);
         \draw [very thick, ->] (2.55,2) -- (2.6,2);
     % horizontal lines      
   \draw [very thick, -] (-4,1) -- (3,1);
   % arrows
         \draw [very thick, ->] (-3.55,1) -- (-3.6,1);
         \draw [very thick, ->] (-2.5,1) -- (-2.45,1);
         \draw [very thick, ->] (-1.55,1) -- (-1.6,1);
         \draw [very thick, ->] (-0.5,1) -- (-0.45,1);
         \draw [very thick, ->] (0.5,1) -- (0.45,1);
         \draw [very thick, ->] (1.55,1) -- (1.6,1);
         \draw [very thick, ->] (2.5,1) -- (2.45,1);
     % horizontal lines      
   \draw [very thick, -] (-4,0) -- (3,0);
   % arrows
        % arrows
        \draw [very thick, ->] (-3.55,0) -- (-3.5,0);
         \draw [very thick, ->] (-2.5,0) -- (-2.55,0);
         \draw [very thick, ->] (-1.55,0) -- (-1.5,0);
         \draw [very thick, ->] (-0.5,0) -- (-0.55,0);
         \draw [very thick, ->] (0.45,0) -- (0.5,0);
         \draw [very thick, ->] (1.5,0) -- (1.45,0);
         \draw [very thick, ->] (2.55,0) -- (2.6,0);
% horizontal lines      
   \draw [very thick, -] (-4,-1) -- (3,-1);
      % arrows
         \draw [very thick, ->] (-3.55,-1) -- (-3.6,-1);
         \draw [very thick, ->] (-2.5,-1) -- (-2.45,-1);
         \draw [very thick, ->] (-1.55,-1) -- (-1.6,-1);
         \draw [very thick, ->] (-0.5,-1) -- (-0.45,-1);
         \draw [very thick, ->] (0.5,-1) -- (0.45,-1);
         \draw [very thick, ->] (1.55,-1) -- (1.6,-1);
         \draw [very thick, ->] (2.5,-1) -- (2.45,-1);
% horizontal lines      
    \draw [very thick, -] (-4,-2) -- (3,-2);
        % arrows
         \draw [very thick, ->] (-3.55,-2) -- (-3.6,-2);
         \draw [very thick, ->] (-2.5,-2) -- (-2.45,-2);
         \draw [very thick, ->] (-1.55,-2) -- (-1.6,-2);
         \draw [very thick, ->] (-0.5,-2) -- (-0.45,-2);
         \draw [very thick, ->] (0.5,-2) -- (0.45,-2);
         \draw [very thick, ->] (1.55,-2) -- (1.6,-2);
         \draw [very thick, ->] (2.5,-2) -- (2.45,-2);
%%%%%%%% The magnetic field %%%%%%%%%%%
  \draw [very thick, ->] (4,-1) -- (4,0);
  \Text(135,0)[]{$\vec B$}; 
   \end{tikzpicture}
\end{center}
\caption{\small The magnetic field $\vec B$ orders the spins on all columns. The rows
have staggered antiferromagnetic order but their first spin is free to chose among the 
two possible orientations. There is a residual entropy $S_{\rm res} \propto \ln N_{\rm columns}$. 
There is staggered two-in two-out  order on each row but no special order between rows.}
\label{fig:tres}
\end{figure}

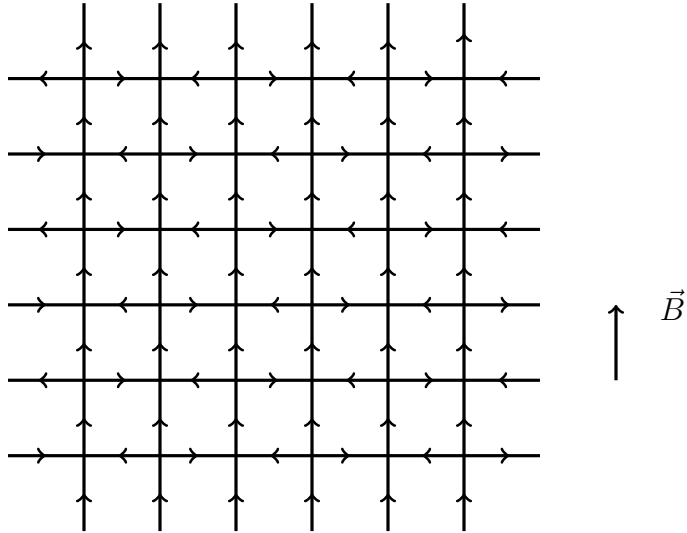
\begin{figure}
\begin{center}
\begin{tikzpicture}
% vertical lines 
   \draw [very thick, -] (-3,-3) -- (-3,4);
% arrows
         \draw [very thick, ->] (-3,-2.55) -- (-3,-2.5);
         \draw [very thick, ->] (-3,-1.55) -- (-3,-1.5);
         \draw [very thick, ->] (-3,-0.55) -- (-3,-0.5);
         \draw [very thick, ->] (-3, 0.45) -- (-3,0.5);
         \draw [very thick, ->] (-3,1.45) -- (-3,1.5);
         \draw [very thick, ->] (-3,2.45) -- (-3,2.5);
         \draw [very thick, ->] (-3,3.45) -- (-3,3.5);
% vertical lines 
    \draw [very thick, -] (-2,-3) -- (-2,4);
% arrows
         \draw [very thick, ->] (-2,-2.55) -- (-2,-2.5);
         \draw [very thick, ->] (-2,-1.55) -- (-2,-1.5);
         \draw [very thick, ->] (-2,-0.55) -- (-2,-0.5);
         \draw [very thick, ->] (-2, 0.45) -- (-2,0.5);
         \draw [very thick, ->] (-2,1.45) -- (-2,1.5);
         \draw [very thick, ->] (-2,2.45) -- (-2,2.5);
         \draw [very thick, ->] (-2,3.45) -- (-2,3.5);
% vertical line
    \draw [very thick, -] (-1,-3) -- (-1,4);
% arrows
        \draw [very thick, ->] (-1,-2.55) -- (-1,-2.5);
         \draw [very thick, ->] (-1,-1.55) -- (-1,-1.5);
         \draw [very thick, ->] (-1,-0.55) -- (-1,-0.5);
         \draw [very thick, ->] (-1, 0.45) -- (-1,0.5);
         \draw [very thick, ->] (-1,1.45) -- (-1,1.5);
         \draw [very thick, ->] (-1,2.45) -- (-1,2.5);
         \draw [very thick, ->] (-1,3.45) -- (-1,3.5);
% vertical line
    \draw [very thick, -] (0,-3) -- (0,4);
% arrows
        \draw [very thick, ->] (0,-2.55) -- (0,-2.5);
         \draw [very thick, ->] (0,-1.55) -- (0,-1.5);
         \draw [very thick, ->] (0,-0.55) -- (0,-0.5);
         \draw [very thick, ->] (0, 0.45) -- (0,0.5);
         \draw [very thick, ->] (0,1.45) -- (0,1.5);
         \draw [very thick, ->] (0,2.45) -- (0,2.5);
         \draw [very thick, ->] (0,3.45) -- (0,3.5);
% vertical line
    \draw [very thick, -] (1,-3) -- (1,4);
% arrows
        \draw [very thick, ->] (1,-2.55) -- (1,-2.5);
         \draw [very thick, ->] (1,-1.55) -- (1,-1.5);
         \draw [very thick, ->] (1,-0.55) -- (1,-0.5);
         \draw [very thick, ->] (1, 0.45) -- (1,0.5);
         \draw [very thick, ->] (1,1.45) -- (1,1.5);
         \draw [very thick, ->] (1,2.45) -- (1,2.5);
         \draw [very thick, ->] (1,3.45) -- (1,3.5);
% vertical line
    \draw [very thick, -] (2,-3) -- (2,4);
% arrows
        \draw [very thick, ->] (2,-2.55) -- (2,-2.5);
         \draw [very thick, ->] (2,-1.55) -- (2,-1.5);
         \draw [very thick, ->] (2,-0.55) -- (2,-0.5);
         \draw [very thick, ->] (2, 0.45) -- (2,0.5);
         \draw [very thick, ->] (2,1.45) -- (2,1.5);
         \draw [very thick, ->] (2,2.45) -- (2,2.5);
         \draw [very thick, ->] (2,3.55) -- (2,3.6);
   % horizontal line 1    
   \draw [very thick, -] (-4,3) -- (3,3);
   % arrows
         \draw [very thick, ->] (-3.55,3) -- (-3.6,3);
         \draw [very thick, ->] (-2.5,3) -- (-2.45,3);
         \draw [very thick, ->] (-1.55,3) -- (-1.6,3);
         \draw [very thick, ->] (-0.5,3) -- (-0.45,3);
         \draw [very thick, ->] (0.5,3) -- (0.45,3);
         \draw [very thick, ->] (1.55,3) -- (1.6,3);
         \draw [very thick, ->] (2.5,3) -- (2.45,3);
% horizontal line 2      
   \draw [very thick, -] (-4,2) -- (3,2);
     % arrows
        \draw [very thick, ->] (-3.55,2) -- (-3.5,2);
         \draw [very thick, ->] (-2.5,2) -- (-2.55,2);
         \draw [very thick, ->] (-1.55,2) -- (-1.5,2);
         \draw [very thick, ->] (-0.5,2) -- (-0.55,2);
         \draw [very thick, ->] (0.45,2) -- (0.5,2);
         \draw [very thick, ->] (1.5,2) -- (1.45,2);
         \draw [very thick, ->] (2.55,2) -- (2.6,2);
     % horizontal line 3      
   \draw [very thick, -] (-4,1) -- (3,1);
   % arrows
         \draw [very thick, ->] (-3.55,1) -- (-3.6,1);
         \draw [very thick, ->] (-2.5,1) -- (-2.45,1);
         \draw [very thick, ->] (-1.55,1) -- (-1.6,1);
         \draw [very thick, ->] (-0.5,1) -- (-0.45,1);
         \draw [very thick, ->] (0.5,1) -- (0.45,1);
         \draw [very thick, ->] (1.55,1) -- (1.6,1);
         \draw [very thick, ->] (2.5,1) -- (2.45,1);
     % horizontal lines  4
   \draw [very thick, -] (-4,0) -- (3,0);
   % arrows
        \draw [very thick, ->] (-3.55,0) -- (-3.5,0);
         \draw [very thick, ->] (-2.5,0) -- (-2.55,0);
         \draw [very thick, ->] (-1.55,0) -- (-1.5,0);
         \draw [very thick, ->] (-0.5,0) -- (-0.55,0);
         \draw [very thick, ->] (0.45,0) -- (0.5,0);
         \draw [very thick, ->] (1.5,0) -- (1.45,0);
         \draw [very thick, ->] (2.55,0) -- (2.6,0);
% horizontal lines  5    
   \draw [very thick, -] (-4,-1) -- (3,-1);
      % arrows
         \draw [very thick, ->] (-3.55,-1) -- (-3.6,-1);
         \draw [very thick, ->] (-2.5,-1) -- (-2.45,-1);
         \draw [very thick, ->] (-1.55,-1) -- (-1.6,-1);
         \draw [very thick, ->] (-0.5,-1) -- (-0.45,-1);
         \draw [very thick, ->] (0.5,-1) -- (0.45,-1);
         \draw [very thick, ->] (1.55,-1) -- (1.6,-1);
         \draw [very thick, ->] (2.5,-1) -- (2.45,-1);
% horizontal lines   6   
   \draw [very thick, -] (-4,-2) -- (3,-2);
   % arrows
        \draw [very thick, ->] (-3.55,-2) -- (-3.5,-2);
         \draw [very thick, ->] (-2.5,-2) -- (-2.55,-2);
         \draw [very thick, ->] (-1.55,-2) -- (-1.5,-2);
         \draw [very thick, ->] (-0.5,-2) -- (-0.55,-2);
         \draw [very thick, ->] (0.45,-2) -- (0.5,-2);
         \draw [very thick, ->] (1.5,-2) -- (1.45,-2);
         \draw [very thick, ->] (2.55,-2) -- (2.6,-2);
%%%%%%%% The magnetic field %%%%%%%%%%%
  \draw [very thick, ->] (4,-1) -- (4,0);
  \Text(135,0)[]{$\vec B$}; 
  %%%%% Circles %%%%%%%%%%%%%%
\end{tikzpicture}
\end{center}
\caption{\small Two special ground states under the field.}
\label{fig:cuatro}
\end{figure}

Take the sixteen vertex model. Under no applied magnetic field 
there is spin reversal symmetry, 
$\omega_1= \omega_2$,
$\omega_3= \omega_4$,
$\omega_5= \omega_6$, 
$\omega_7= \omega_8$,
$\omega_9= \omega_{10}$,
$\omega_{11}= \omega_{12}$,
$\omega_{13}= \omega_{14}$,
$\omega_{15}= \omega_{16}$.
Assume that the vertex energies are ordered 
according to $\epsilon_c < \epsilon_e < \epsilon_a =\epsilon_b < \epsilon_d$, a quite 
unusual hierarchy since the vertices with three-in one-out and three-out one-in arrows 
are usually considered to be defects with relative high energy. (However, there should be tricks to 
realise this ordering in the laboratory.)
As the anti-ferroelectric vertices are the ones with the lowest energy, the 
ground state has staggered order of $c$ vertices.
The magnetic field lifts the degeneracy between the energies of the 
vertices that have two vertical arrows pointing in the same direction,
that is to say, the vertices labeled 1 and 2, 2 and 3 and between some 
of the $e$ vertices, see Fig.~\ref{fig:uno}. For a sufficiently strong magnetic field, one can in this 
way render the energy of the $e$ vertices with the two vertical arrows aligned along 
the magnetic field the lowest one (vertices labeled 14 and 15 in Fig.~\ref{fig:uno}). In Fig.~\ref{fig:tres}
we show one of the ground states among the $2^L$ with $L$ the number of lines, possible ones under these conditions. 
While all arrows on vertical edges are aligned with the magnetic field,  perpendicular 
arrows are free to point in any of the two directions, leading to these degeneracy.
One notices that among all these possible ground states, two of them
are completely anti-ferromagnetically ordered along rows and between rows 
and are shown in Fig.~\ref{fig:cuatro}. For magnetic fields
that are just above the threshold value at which the energy hierarchy is modified to $\epsilon_e<\epsilon_c$, these are the 
ground states with the largest number of low-energy excitations. They correspond to flipping alternating  
vertical arrows against the magnetic field, but with low energetic cost. This feature produces an effective
anti-ferromagnetic interaction between the lines that fully orders the system antiferromagnetically at low temperatures.
The careful analysis of the finite size effects shows that the low-temperature and large system 
size limits do not commute. This is intimately related to the fact that ordering at infinitesimal temperature 
appears as a first order phase transition. The non-trivial dependence on system size could 
be exploited to detect ObD experimentally, as explained in Ref.~\cite{Guruciaga16}.

\vspace{0.25cm}
\noindent
{\bf Dynamics}
\vspace{0.25cm}

Several kinds of stochastic dynamic rules have been proposed to study different aspects of the dynamics and statics of vertex models.

In the pure vertex model context, the main interest has been to study 
the Kosterlitz-Thouless phase transition 
between disordered and anti-ferromagnetic phases in the six vertex model  with numerical methods.  The 
elementary moves can not violate the strict two-in two-out rule, that is equivalent to a `divergence free' condition in an
analogy of the sequence of arrows with magnetic lines, that should form, therefore, closed loops.
{\it Loop algorithms}, in which the orientation of closed loops is reversed using a local
stochastic decision that respects detailed balance, 
were specifically developed to beat critical slowing down~\cite{Evertz93}. Improvements of this method 
to study other features of vertex models and their quantum spin-chain equivalents are still now being proposed.

As soon as the strict ice rule is broken, other algorithms can be used to mimic the dynamics of these 
systems and at long times sample their asymptotic states.

If the aim is to use the vertex models to describe the behaviour of real or artificial materials, details on their 
actual dynamics have to be taken into account when defining the microscopic updates.
In their final state  the magnetic dots in artificial spin-ice are small enough to be single-domain, 
but large enough to be athermal.  In consequence, the blocked configurations reached, for example, 
with rotating magnetic field protocols,  are the result of athermal non-equilibrium 
dynamics with some similarity to the shaking  of granular materials.
In several works emphasis was put on the study of the statistical properties of the steady state reached with 
these and other athermal evolutions~\cite{Nisoli10,Nisoli12}.

In other experimental protocols, the evolution of the spin-ice configuration is thermal~\cite{Wang2006,Morgan10}
since the magnetic elements can flip during their formation by deposition before a critical size is reached.
The domain structure and formation under thermal fluctuations 
was studied in~\cite{Budrikis12b} using the iteration of a mean-field equation 
for the local magnetizations based on a point dipole approximation, 
and Monte Carlo simulations at the same level of approximation. 

\vspace{0.5cm}

\begin{figure}[h]
\centerline{
\includegraphics[scale=0.6]{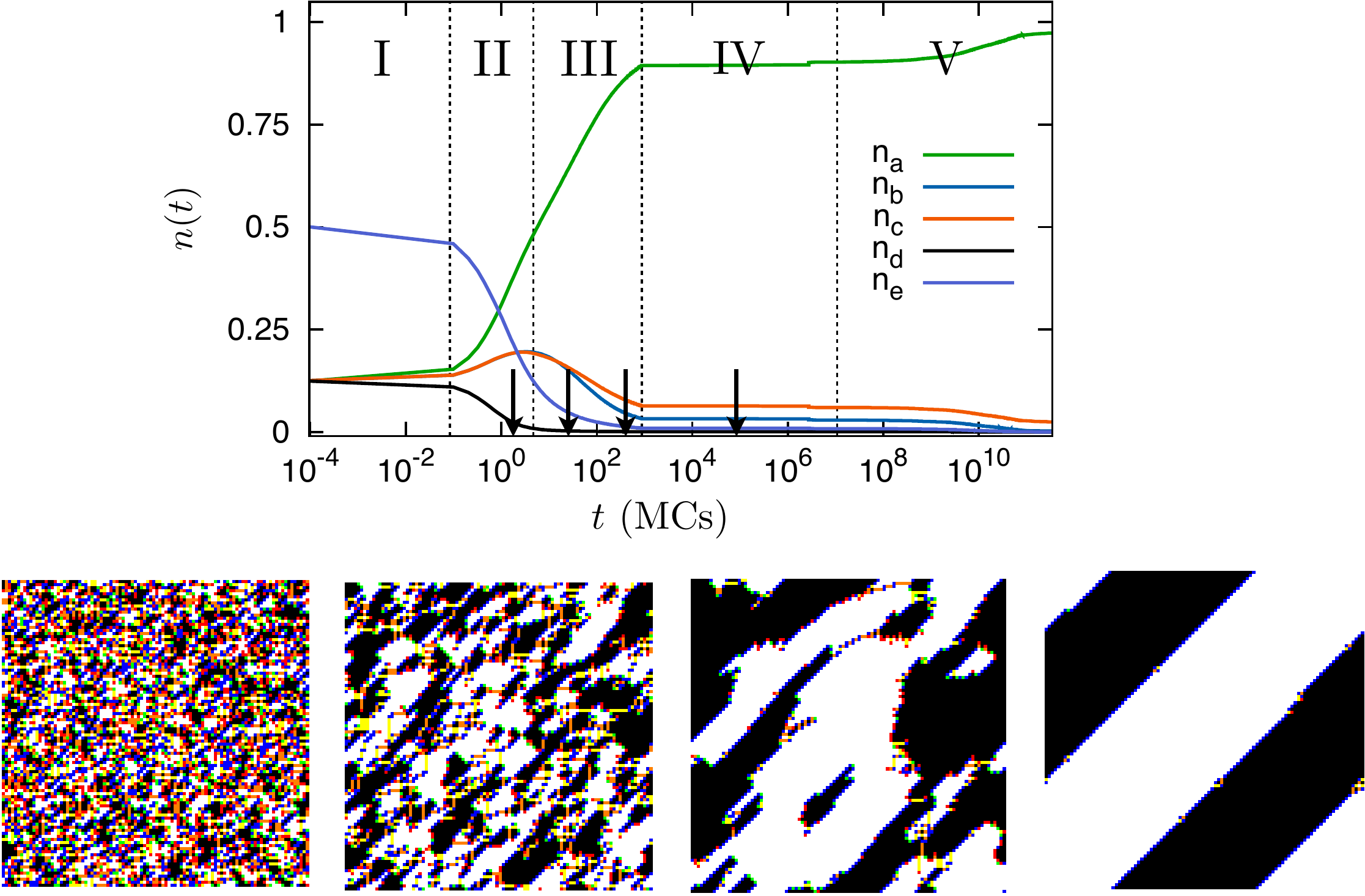}
}
\vspace{0.25cm}
\caption{\small (Colour online.) Progressive ferromagnetic ordering after a quench from a disordered state at $t=0$. 
The time-evolution of the density of vertices  for $a = 5$, $b = 1$, and $d = e^2 = 10^{-10}$, in a system with linear size $L = 100$. 
The data are averaged over 300 samples. The snapshots are typical
configurations at the instants indicated by the arrows.Black and white points are vertices 1 and 2.  This figure is taken from Ref.~\cite{Levis2013b}.
The four regimes labeled I, II, III, IV, and V in the figure are discussed in this reference.}
\label{fig:domain-growth}
\end{figure}

In a couple of papers we studied the dynamics after thermal quenches in the 2d square lattice 
spin-ice model built as a stochastic extension of the vertex models~\cite{Levis2012,Levis2013b}. 
We mimicked the effect of thermal fluctuations in spin-ice
samples by coupling the model to an environment and allowing for local single spin flips determined by the heat-bath rule. 
Local moves that break the spin-ice rule are not forbidden and we 
therefore allow for thermally activated creation of defects in the form of three-in one-out, four-in none-out and their spin-reversed configurations. 
These dynamics do not conserve any of the various order parameters 
and are ergodic for both fixed and periodic boundary conditions. With these ingredients we 
established a Monte Carlo algorithm and we defined the unit of time as a Monte Carlo sweep (MCs). In systems with frustration, 
computer time is wasted by the large 
rejection of blindly proposed updates. 
To avoid this problem we used a rejection-free continuous-time Monte Carlo (CTMC) algorithm that allows for thermally activated creation of defects. 
The longest time reached with this method, once translated in terms of usual MC sweeps, is of the order of $10^{25}$ MCs, a scale practically
unreachable with usual algorithms. 

The CTMC dynamics allowed us to identify the equilibrium phase diagram and to analyse different dynamic regimes. 
Our dynamic results are manifold~\cite{Levis2012,Levis2013b}.
We reproduced known facts of the dynamics of spin-ice samples and we derived a large number of new results that 
should be realized experimentally. 
After a quench to sets of parameters in the disordered phase the system eventually equilibrates but it does in several different time-scales since the systems get 
blocked in long-lived metastable states with a large density of defects.  
Reaction-diffusion arguments~\cite{Castelnovo} were used to understand why these long-lived states exist, although in our model there are no long-range interactions. 
After quenches into the two kinds of ordered phases
the interactions between the spins, mediated by the choice of vertex weights, create ordered domains of ferromagnetic or anti-ferromagnetic kind. 
We proved that the ordering dynamics conforms to the domain-growth scaling picture.
The quantitative characterization of order-growth is given by two growing lengths extracted from correlation
functions along orthogonal directions, $\ell_\parallel(t)$ and  $\ell_\perp(t)$ that, numerically, are both compatible with $t^{1/2}$
though with different pre-factors. In Fig.~\ref{fig:domain-growth} we show an initial configuration with random choice of the vertices and two subsequent snapshots after
having quenched the system into its ferromagnetic phase ($a=5$, $b=1$, $c=1$, $d=e^2 =10^{-1}$). The plots are self-explanatory, with stripes of 
ferromagnetic domains of reversed type shown in black and white. The upper panel displays the time dependence of the density of vertices of different type in the
course of time. The four arrows indicate the instants at which the snapshots were taken. For more details see Refs.~\cite{Levis2012,Levis2013b}.

\section{Conclusions}

We have visited the phase space and real space properties of 2D artificial spin-ice samples as uncovered by their 
study using vertex models. Curiously enough, the analysis of these models with an {\it a priori} crude approximation 
as the one accessed with cluster variational or Bethe-Peierls methods provides many interesting results. 

Dipolar interactions are definitely present in experimental samples and they may alter some of the results presented in this paper. How they may do 
is definitely a very interesting question that deserves to be studied carefully. For instance, one could imagine that the sharp arctic curves may become smoother crossovers 
or that the order-by-disorder phenomenon will have to be searched in samples in which the dipolar interactions are subdominant.

We close by insisting upon the fact that, although vertex
models avoid all the complications of (long-range) dipolar interactions, they are a very convenient schematic framework to study artificial spin-ice samples
from a theoretic perspective, and they are sufficiently rich to have attired and continue to attract the attention of a large number of theoreticians
including myself.

\vspace{0.5cm}

\noindent
{\bf Acknowledgements.}
The research described in this note was carried on between 2011 and 2016 in collaboration with
 R. A. Borzi, M. V. Ferreyra, L. Foini, G. Gonnella, S. A. Grigera, P. C. Guruciaga, D. Levis, A. Pelizzola, J. Restrepo and M. Tarzia, 
all of whom I warmly thank. I also want to thank P. Holdsworth and L. Jaubert for very useful discussions and suggestions.
I acknowledge financial support from ECOS-Sud A14E01, 
PICS 506691
(CNRS-CONICET Argentina) and NSF under Grant No. PHY11-25915.
I am a member of Institut Universitaire de France.

\bibliographystyle{phaip}
\bibliography{MC-Bethe,biblioObD}

\end{document}